\documentclass{aa}
\usepackage{psfig}

\newcommand{\flux}{erg s$^{-1}$ cm$^{-2}$}
\newcommand{\tcdm}{$\tau$CDM\ }
\newcommand{\lcdm}{$\Lambda$CDM\ }
\newcommand{\ocdm}{OCDM\ }
\newcommand{\logns}{$\log N - \log S$}

\begin{document}

\title{Cosmology with Galaxy Clusters in the XMM Large-Scale Structure
  Survey}

\author{Alexandre Refregier$^{1}$, Ivan Valtchanov$^{2}$,
\& Marguerite Pierre$^{2}$}
\institute{$^{1}$ Institute of Astronomy, Madingley Road, Cambridge
CB3 OHA, UK; ar@ast.cam.ac.uk\\
$^{2}$ Service d'Astrophysique, Bat. 709, CEA Saclay, F-91191,
Gif-sur-Yvette, France; ivaltchanov, mpierre@cea.fr}

\titlerunning{Clusters in the XMM-LSS}
\authorrunning{Refregier et al}

\date{Received, ~~~, Accepted, ~~~}

\markboth{Clusters in the XMM-LSS}{Refregier et al.}

\abstract{The upcoming XMM Large Scale Structure Survey (XMM-LSS) will
ultimately provide a unique mapping of the distribution of X-ray
sources in a contiguous 64 deg$^{2}$ region. In particular, it will
provide the 3-dimensional location of about 900 galaxy clusters out to
a redshift of about 1. We study the prospects that this cluster
catalogue offers for measuring cosmological parameters. We use the
Press-Schechter formalism to predict the counts of clusters and their
X-ray properties in several CDM models. We compute the detection
efficiency of clusters, using realistic simulations of XMM X-ray
images, and study how it differs from a conventional flux limit. We
compute the expected correlation function of clusters using the
extended halo model, and show that it is expected to evolve very
little out to $z\simeq 2$, once the selection function of the survey
is taken into account. The shape and the amplitude of the correlation
function can be used to brake degeneracies present when cluster counts
alone are considered. Ignoring systematic uncertainties, the
combination of cluster counts evolution and of the correlation
function yields measurements of $\Omega_{m}$, $\sigma_{8}$ and
$\Gamma$ with a precision of about 15\%, 10\% and 35\%, respectively,
in a \lcdm model. This combination will also provide a consistency
check for the \lcdm model, and a discrimination between this model and
the \ocdm model. The XMM-LSS will therefore provide important
constraints on cosmological parameters, complementing that from other
methods such as the Cosmic Microwave Background. We discuss how
these constraints are affected by instrumental systematics and by the
uncertainties in the scaling relations of clusters.  
 }

\maketitle

\section{Introduction}
Clusters of galaxies are the most massive bound objects in the
Universe and provide a powerful cosmological probe (see e.g. Borgani
\& Guzzo \cite{bor01b} for a review). In particular, the number counts
of clusters and its evolution yield a robust measure of both the
amplitude of the matter power spectrum and of the geometry of the
universe (e.g. Oukbir \& Blanchard \cite{ouk97}; Eke et al.
\cite{eke98}; Viana \& Liddle \cite{via99}; Kitiyama \& Suto
\cite{kit97}).  The spatial correlation function of clusters
quantifies the clustering of these objects and yields complementary
constraints on cosmology (e.g. Mo, Jing \& White \cite{mo96b}; Suto et
al. \cite{sut00}; Robinson \cite{rob00}; Moscardini et
al. \cite{mos00}; Collins et al. \cite{coll00}).

In this paper, we explore the prospects of measuring cosmological
parameters with the upcoming XMM Large Scale Structure Survey
(XMM-LSS; Pierre \cite{mp00}). This survey consists of 10 ksec
exposures of an $8\times 8$ deg$^{2}$ region with the XMM-Newton
observatory, along with an extensive follow-up programme in the
optical, IR and radio bands. In particular, it will provide the
3-dimensional location of about 900 clusters out to a redshift of
about 1.  Thanks to its uniform sensitivity across a contiguous
region, this survey thus provides a unique database to measure the
evolution of both the number counts and the correlation function of
clusters.

To study how the clusters found in XMM-LSS can constrain cosmological
models, we use the Press \& Schechter (\cite{ps74}) formalism to
predict the expected cluster counts in the survey.  This is done using
the selection function of the survey derived from detailed simulations
of cluster detections in XMM-Newton images (see Valtchanov, Pierre \&
Gastaud \cite{val01}, VPG).  Using the Mo \& White (\cite{mo96a})
formalism, we compute the expected spatial correlation function for
the detected clusters.  We then study how the cluster counts and
correlation function, taken together, constrain cosmological
parameters.  

Our analysis extends the work of Moscardini et al.  
(\cite{mos00}) who considered the expected cluster counts and
correlation function for XMM-LSS. They however assumed a simple flux
limit, rather than the more realistic selection function which we
consider. In addition, they did not compute the cosmological
constraints from a joint measurement of the cluster counts and
correlation function with XMM-LSS. Our results also complement the
analysis of Romer et al. (\cite{rom01}) who studied the cosmological
dependence of cluster counts in the planned Serendipitous XMM Cluster
Survey. They are also related to the work of Haiman al. (\cite{hai01})
and Holder et al. (\cite{hol01}) who studied the cosmological
constraints which can be derived from the evolution of cluster counts
with a dedicated wide-angle X-ray mission.

The paper is organized as follows. In \S\ref{XMM-LSS} we summarize the
characteristics of the XMM-LSS. In \S\ref{simulations}, we describe
the simulations for cluster detection and derive the cluster selection
function. In \S\ref{counts} we compute the expected cluster counts
using the selection function combined with the Press-Schechter
formalism. In \S\ref{correlation} we compute the correlation function
for this cluster sample and show how it constrains cosmological
parameters. The effects of systematic uncertainties on these
constraints are discussed in \S\ref{systematics}. Our conclusions are
summarized in \S\ref{conclusion}.

\section{The XMM Large Scale Structure Survey}
\label{XMM-LSS} The XMM-LSS Survey is a unique medium-deep cluster
survey combining X-ray observations with an extensive optical, IR, and
radio follow-up programme (Pierre \cite{mp00}). The survey geometry -
coverage and depth - was chosen to allow the measurement of the
cluster two-point correlation function, with better than 15\% error on
the correlation length, in two redshift intervals between $z=0$ and
$1$.

The position of the $8\times 8$ deg$^{2}$ surveyed area on the sky
($\alpha = 2^{h}20^{m}$, $\delta = -5^{o}$) is at a sufficiently high
galactic latitude ($\approx -60^{o}$) in a region of moderate galactic
absorption and without known bright X-ray sources. It will be covered
by $24 \times 24$ partially overlapping XMM pointings with individual
exposure times of 10 ks, reaching a sensitivity of $3 \times 10^{-15}$
\flux\ in the $[0.5-2]$ keV band for point sources, or of about $5
\times 10^{-15}$ \flux\ for cluster-like extended sources. Down to
this limit, some 300 objects (mainly QSOs) per deg$^{2}$ are expected
according to the latest deep surveys (e.g. Hasinger et al.
\cite{has01}, Giacconi et al. \cite{gia01}), and a total of about
10-15 clusters per deg$^{2}$ out to $z \simeq 1$. The survey is also
well suited to probe the existence of massive clusters within the
important $1 < z < 2$ redshift range. At the time of writing, 6
deg$^{2}$ with XMM have been allocated in guaranteed time and guest
observer time. The rest of the survey is subject to an ongoing
application and reviewing process.

An extensive multi-wavelength follow-up programme has been undertaken
by the XMM-LSS consortium\footnote{Official web page of the
consortium: \\http://vela.astro.ulg.ac.be/themes/spatial/xmm/LSS/}.
Special care is given to the optical identification of the X-ray
sources: deep multi-color imaging of the entire region will be
performed by the Canada-France-Hawaii Legacy Survey\footnote{
http://cdsweb.u-strasbg.fr:2001/Instruments/Imaging/ Megacam} and
subsequent redshift measurements by the VIRMOS/VLT instrument and
other large telescopes to which the consortium has access. The main
priorities are: (1) identification and redshift measurement of all
X-ray clusters between $ 0 < z < 1 $, (2) NIR observations of distant
($ z > 1 $) cluster candidates and, subsequently, determination of
their redshift, (3) serendipitous spectroscopic observations of the
X-ray QSOs, in order to study their clustering properties within the deep
potential-well network traced by the clusters.

\section{Simulations}
\label{simulations} 
In order to estimate the detection probability of clusters in the
survey, we performed a series of X-ray image simulations. While a
detailed description can be found in VPG, we first review here the
main features of the simulations. We then show how they can be used to
derive the selection function for the survey.

\subsection{Cluster Detection in Simulated Images}

The simulations reproduce the main characteristics of the XMM-EPIC
instruments, such as the Point-Spread Function (assuming circular
symmetry for the PSF shape; see VPG) and vignetting as a function of
energy and off-axis angle. These were parametrized using the latest
available on-flight calibrations. For the diffuse and particle
background, we have used the data from Watson et al.  (\cite{wat01}).

The point-like sources were laid at random inside the field-of-view.
Their fluxes were drawn from the $\log N - \log S$ data in Lockman
Hole (Hasinger et al.  \cite{has01}) and Chandra deep field south
(Giacconi et al.  \cite{gia01}). Their spectrum was taken to be a
power law with photon index 2.

Clusters of galaxies were modeled as spherically symmetric objects
assuming a $\beta$-profile (e.g.  Cavaliere \& Fusco-Femiano
\cite{cff76}) with fixed core radius $r_c=125\,h^{-1}$ kpc and slope
$\beta=0.75$ (see \S\ref{systematics} for a discussion of the
effect of a varying core radius on our predictions). A thermal plasma
spectrum (Raymond \& Smith \cite{rs77}) was assumed. The spectrum was
normalized using the non-evolving luminosity-temperature ($L-T$)
relation of Arnaud \& Evrard (\cite{arn99}). When generalized to
arbitrary cosmological model, it is given by
\begin{equation}
\label{eq:l_t}
L = 2.87 \times 10^{44}\left(\frac{T}{6{\rm keV}} \right)^{2.88}
\left( \frac{D_{L}}{D_{L,{\rm EdS}}} \right)^{2} h^{-2}
{\rm erg}~{\rm s}^{-1},
\end{equation}
where $L$ is the bolometric luminosity, $T$ is the X-ray temperature,
and $D_{L}$ and $D_{L,{\rm EdS}}$ are the luminosity distances in the
desired and Einstein-de Sitter cosmological models, respectively.

We took the neutral hydrogen column density to be $N_H=5\times
10^{20}$ cm$^{-2}$ and element abundances $Z=0.3Z_{\odot}$. Using
XSPEC (Arnaud \cite{xspec}), we calculated the total expected count
rates for the extended and point-like sources for the three XMM EPIC
instruments\footnote{See e.g. the XMM-Newton User's Handbook:\\
  http://xmm.vilspa.esa.es/user/uhb/xmm\_uhb.html} with thin filters
in [0.5-2] keV energy band, for an integration time of 10 ks.

\begin{figure*}
  \centerline{
    \psfig{figure=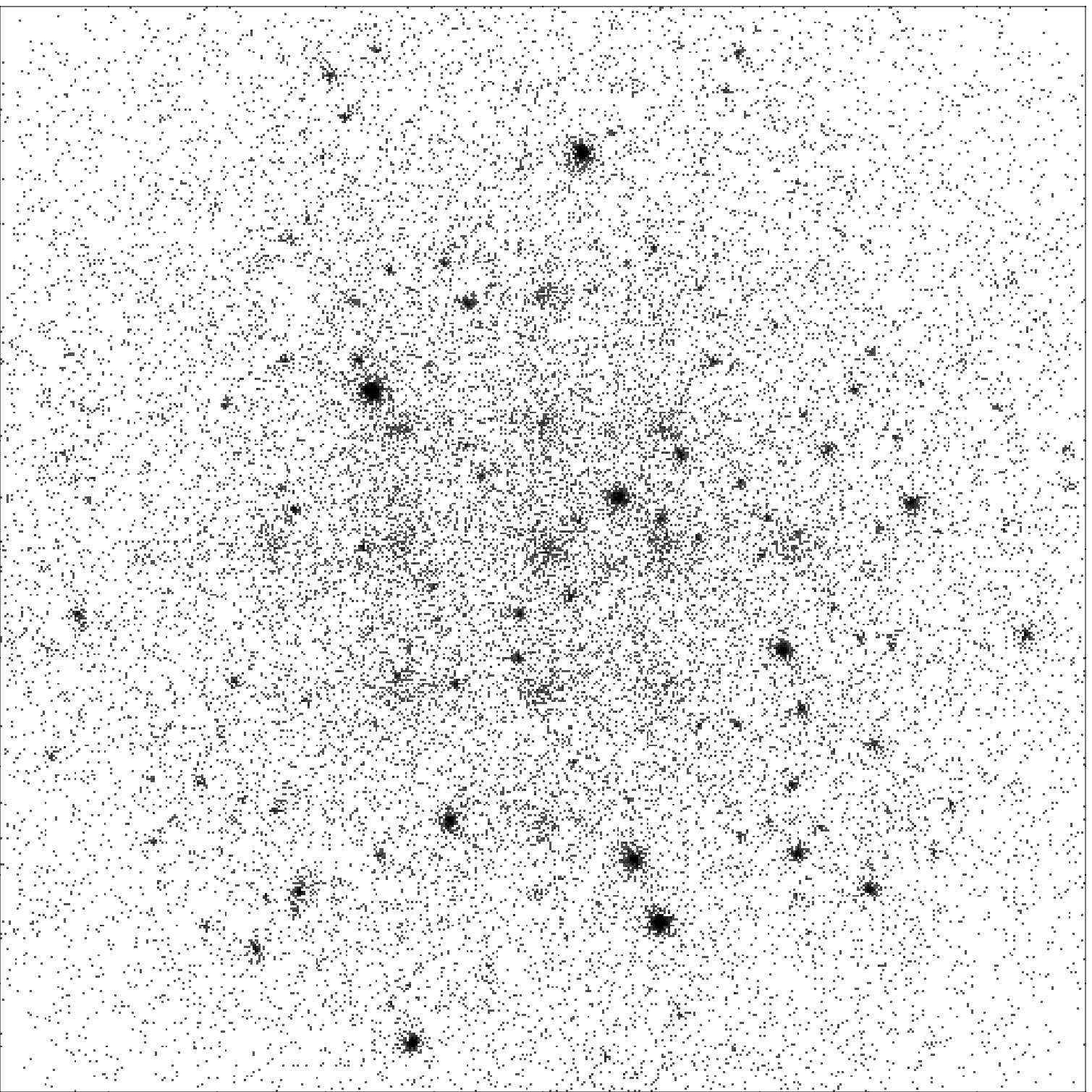,width=8cm} \hfill
    \psfig{figure=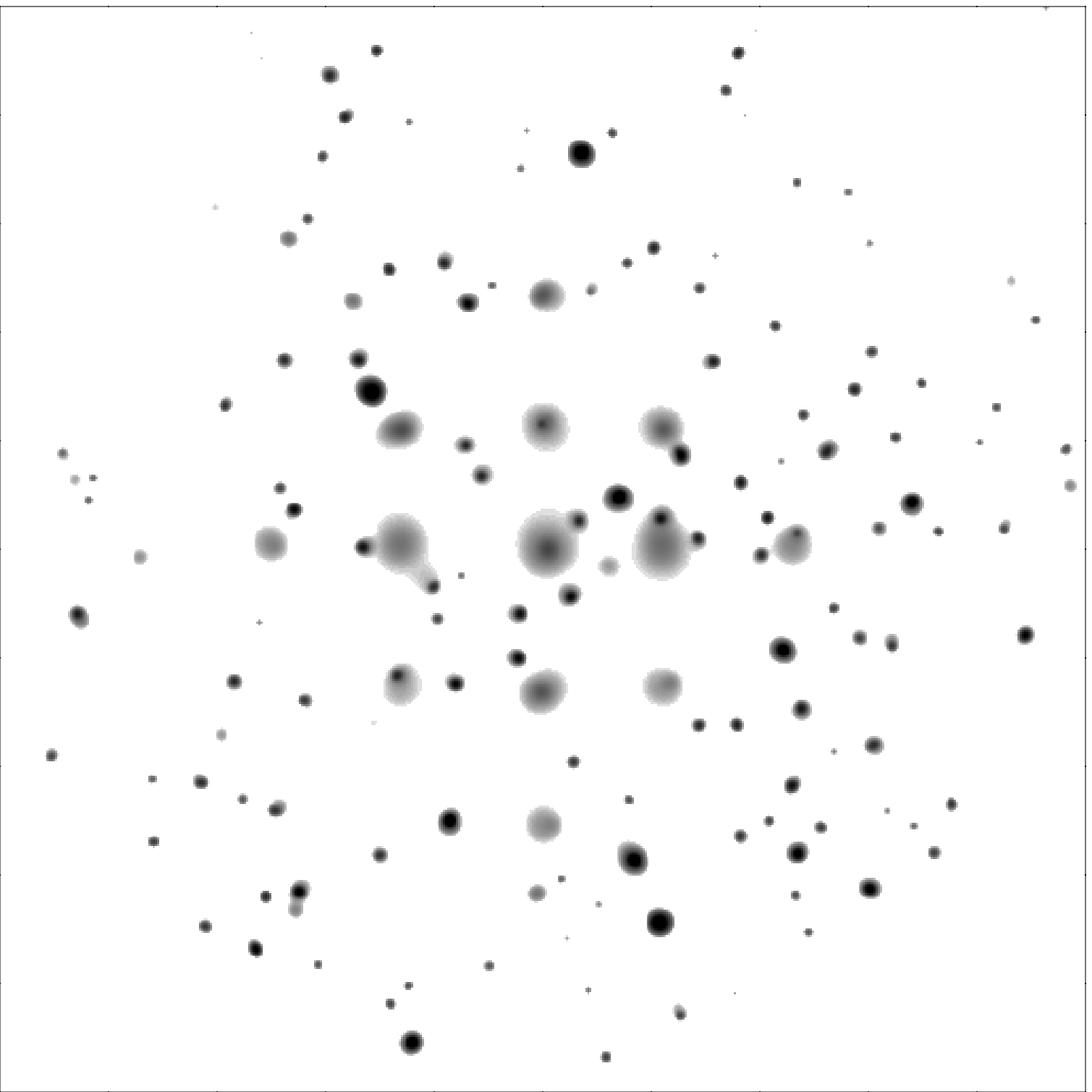,width=8cm}}
  \caption{Simulated extragalactic XMM-LSS raw image (left) and the
    corresponding wavelet filtered image with Poisson noise model and
    $10^{-4}$ ($\sim 4\sigma$) significance (Starck \& Pierre 1998).
    The energy band is $[0.5-2]$ keV, the exposure time 10ks, and the
    three XMM instruments are added together (MOS1, MOS2 and pn).
    Point-like sources follow the observed \logns\ relation and the
    extended sources (clearly visible on the wavelet filtered image)
    are clusters of galaxies with $T_X = 3$ keV at $z=1$ (see text for
    details).}
  \label{fig:sim1}
\end{figure*}

\begin{figure*}
  \centerline{
    \psfig{figure=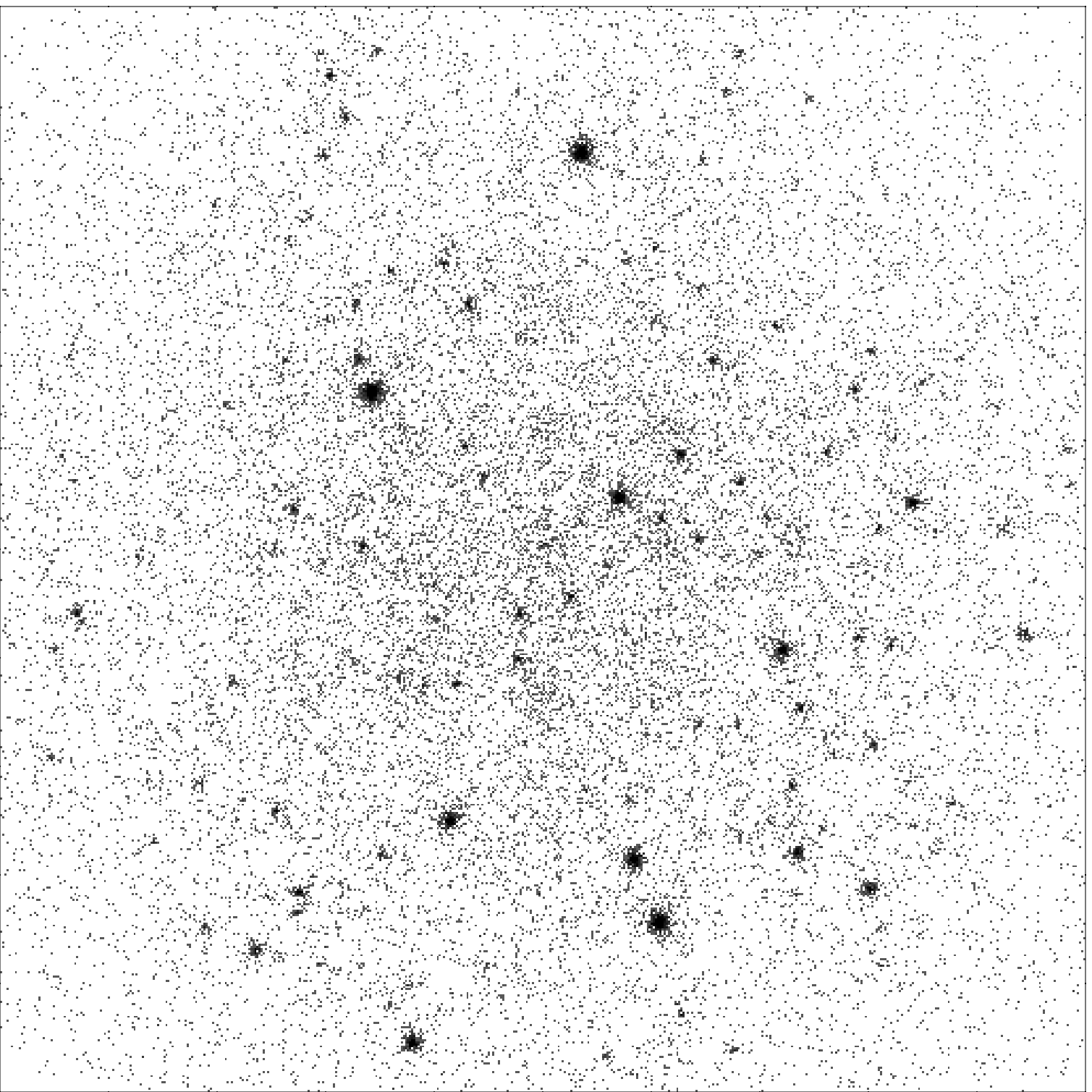,width=8cm} \hfill
    \psfig{figure=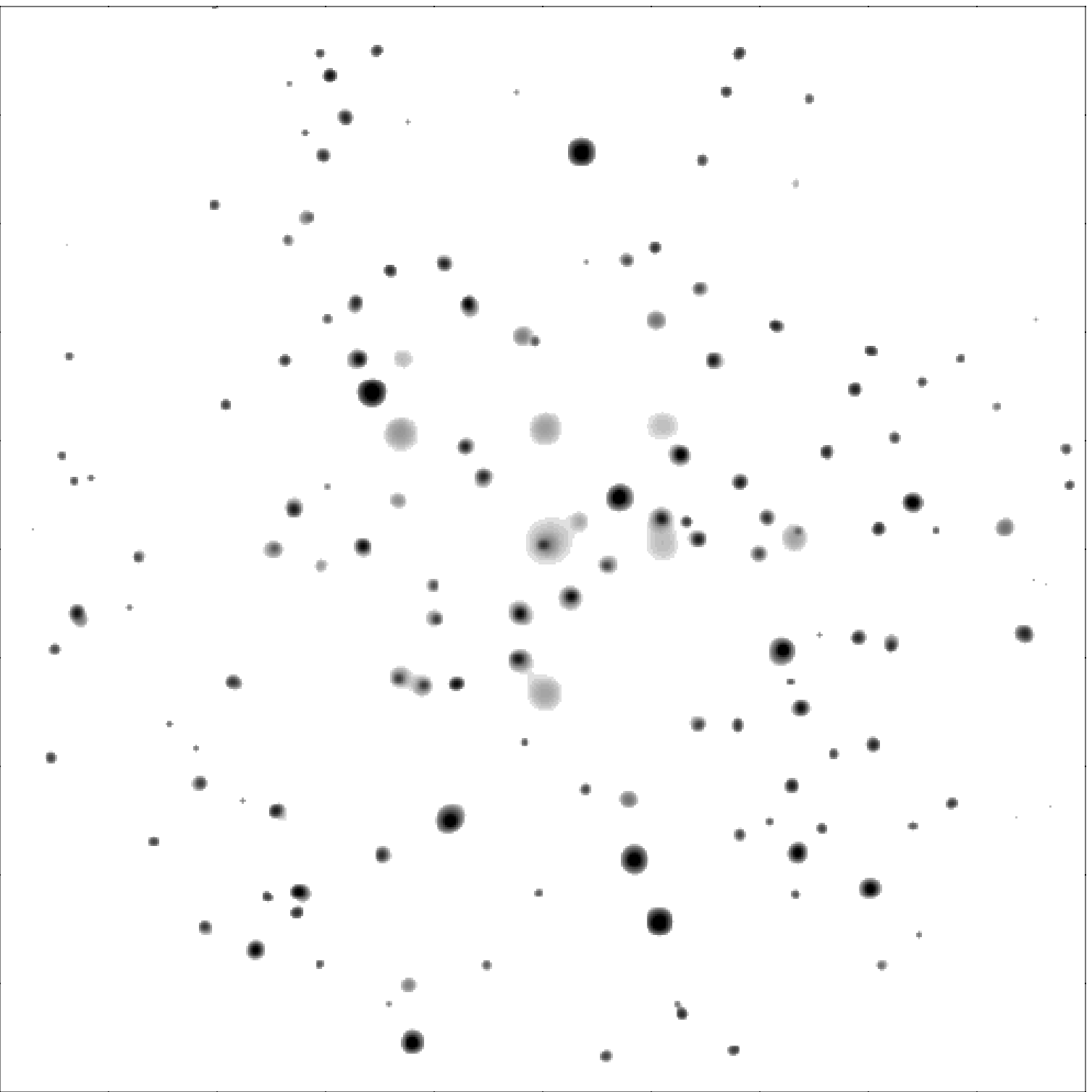,width=8cm} }
  \caption{Same as Fig.~\ref{fig:sim1} but with clusters at $z=1.5$.}
  \label{fig:sim2}
\end{figure*}

For a given temperature and redshift, we placed 13 clusters with
centers on a grid inside the inner $10\arcmin$ radius of the
field-of-view. This number was chosen to maximally fill this region
while avoiding overlaps. The grid pattern can be seen in the
right-hand panel of Figure~\ref{fig:sim1}. The detection was
performed in the same way as in VPG, i.e. using multi-scale (wavelet)
filtering assuming Poisson noise statistics (Starck \& Pierre
\cite{sp98}), followed by {\tt SExtractor} (Bertin \& Arnouts
\cite{sex}) detection and classification. As was pointed out in VPG,
this is currently the most suitable method to detect, characterize and
classify extended sources in XMM images. The raw photon and wavelet
filtered images for clusters with $T = 3$ keV at a redshifts of $z=1$
and 1.5 are shown in Figs.~\ref{fig:sim1} and \ref{fig:sim2}.

To cross-identify the detected objects with the input clusters we have
used a searching radius of $12\arcsec$. If a correspondence is found,
we perform a classification based on the half-light radius and the
stellarity index, to determine whether the object is extended. (For
the choice of the searching radius and the classification criteria,
see VPG). A cluster is finally considered to be detected if the
positional {\em and} the classification criteria are obeyed. This
procedure is close to the planned analysis of the incoming XMM data,
which will make use of the multi-color optical data to confirm
the existence of the X-ray cluster candidates.

\subsection{Selection Function}
\label{selection_function}
The selection function is a convenient way of quantifying the
detection sensitivity of a survey. It was computed for a number of
existing and future cluster surveys (eg. Romer et al. \cite{rom01}
for the XMM serendipitous Cluster Survey; Henry et
al. \cite{hen01} for the NEP ROSAT survey; Adami et al. \cite{ada00}
for the SHARC survey) using both analytical estimates and image
simulations.

\begin{figure}
\centerline{
\psfig{figure=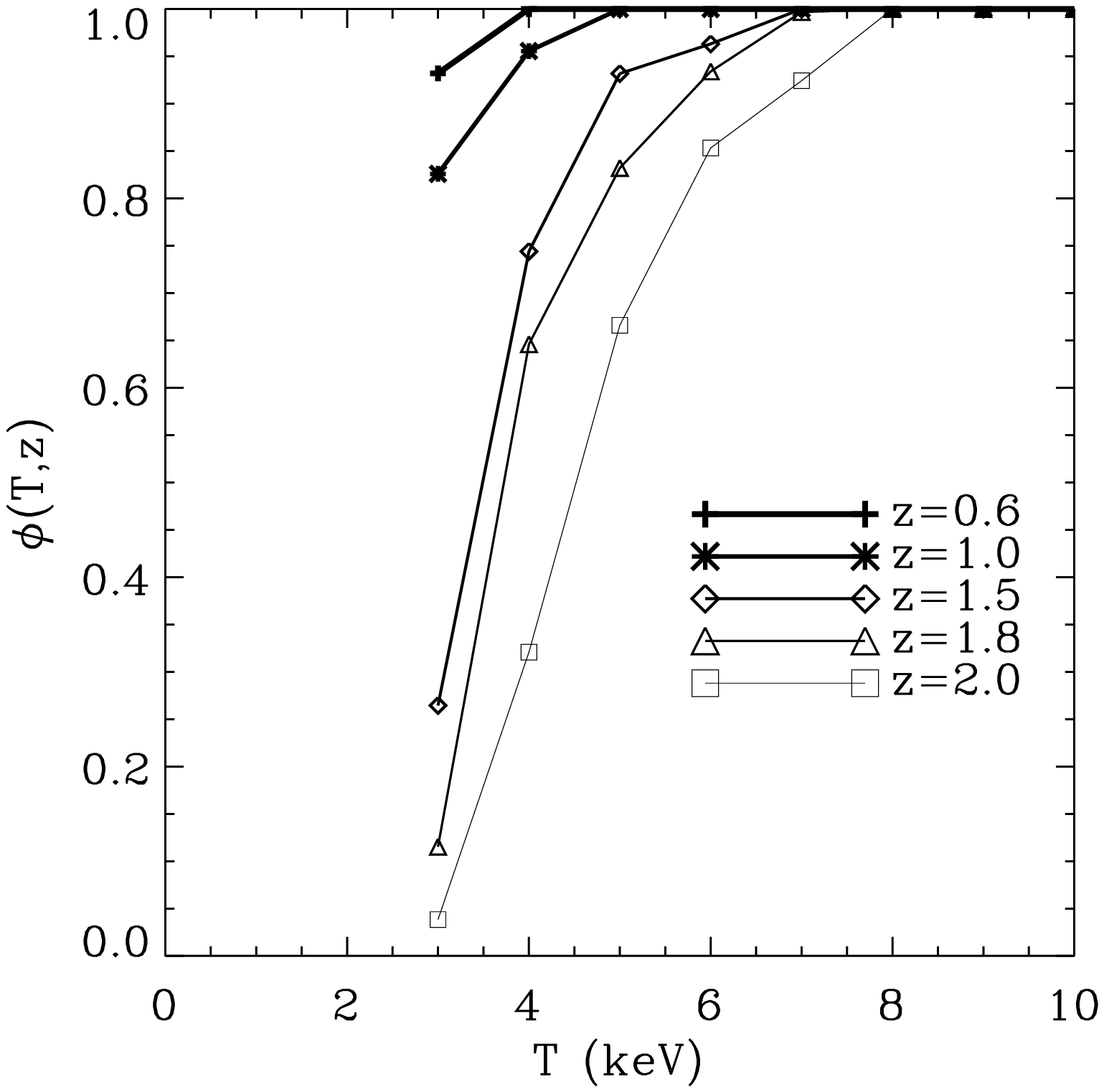,width=8cm}}
\caption{\label{fig:phi_tz}Selection function $\phi(T,z)$ for
  \lcdm model derived from the simulations. This function is
  the probability that a cluster with temperature $T$ and redshift $z$
  is detected and classified as an extended object in the survey
  catalogue.}
\end{figure}

Here, we estimate the selection function of the XMM-LSS, by performing
a set of 10 simulations for a set of temperatures ($T=3,4,5,6,7,8,9$
keV) and redshifts ($z=0.6,1.0,1.5,1.8,2.0$). For each value of $T$
and $z$, the selection $\phi(T,z)$ of clusters for the survey was then
calculated by comparing the number of detections (and correct
classifications) $N_{\rm det}(T,z)$ to the number of input clusters
$N_{\rm in}(T,z)$, so that
\begin{equation}
\label{eq:phi}
\phi(T,z) = N_{\rm det}(T,z)/N_{\rm in}(T,z).
\end{equation}
Fig.~\ref{fig:phi_tz} shows the resulting selection function, which
gives the probability for a cluster with temperature $T$ and redshift
$z$ to be detected and classified as an extended object in the survey
catalogue.  Approximately, 90\% of all clusters with $T > 3$ keV are
detectable out to $z \sim 0.6$. The selection function is close to
about 1 for $T>2$ keV and $z<0.5$ (not shown). Since we are only
interested in clusters (and groups), we set the selection function to
0 for $T<2$ keV at all redshifts.  This corresponds to a minimum
bolometric luminosity of about $1.2\times 10^{43}$ $h^{-2}$ erg
s$^{-1}$ (see Eq.~[\ref{eq:l_t}]) and can thus be easily implemented
in practice. As can be seen on the figure, low temperature clusters
become progressively harder to detect as the redshift increases
(compare Figs.  \ref{fig:sim1} and \ref{fig:sim2}). For example, at
$z=2$ only clusters with $T>6$ keV yield a completeness better than
90\%.

Our results are consistent with that of Romer et
al. (\cite{rom01}; figure~5) who found that essentially all clusters
with with $T>2$ keV ($T>4$ keV) and $z \la 0.6$ ($z \la 1.0$) can be
detected in a survey with a sensitivity comparable to that of
XMM-LSS. Note that our selection function corresponds to the average
sensitivity over the inner 13 arcmin of a single XMM-EPIC
pointing. The effects of sensitivity variations produced by vignetting
and overlapping pointings are discussed in \S\ref{systematics}.

\begin{figure}
\centerline{
\psfig{figure=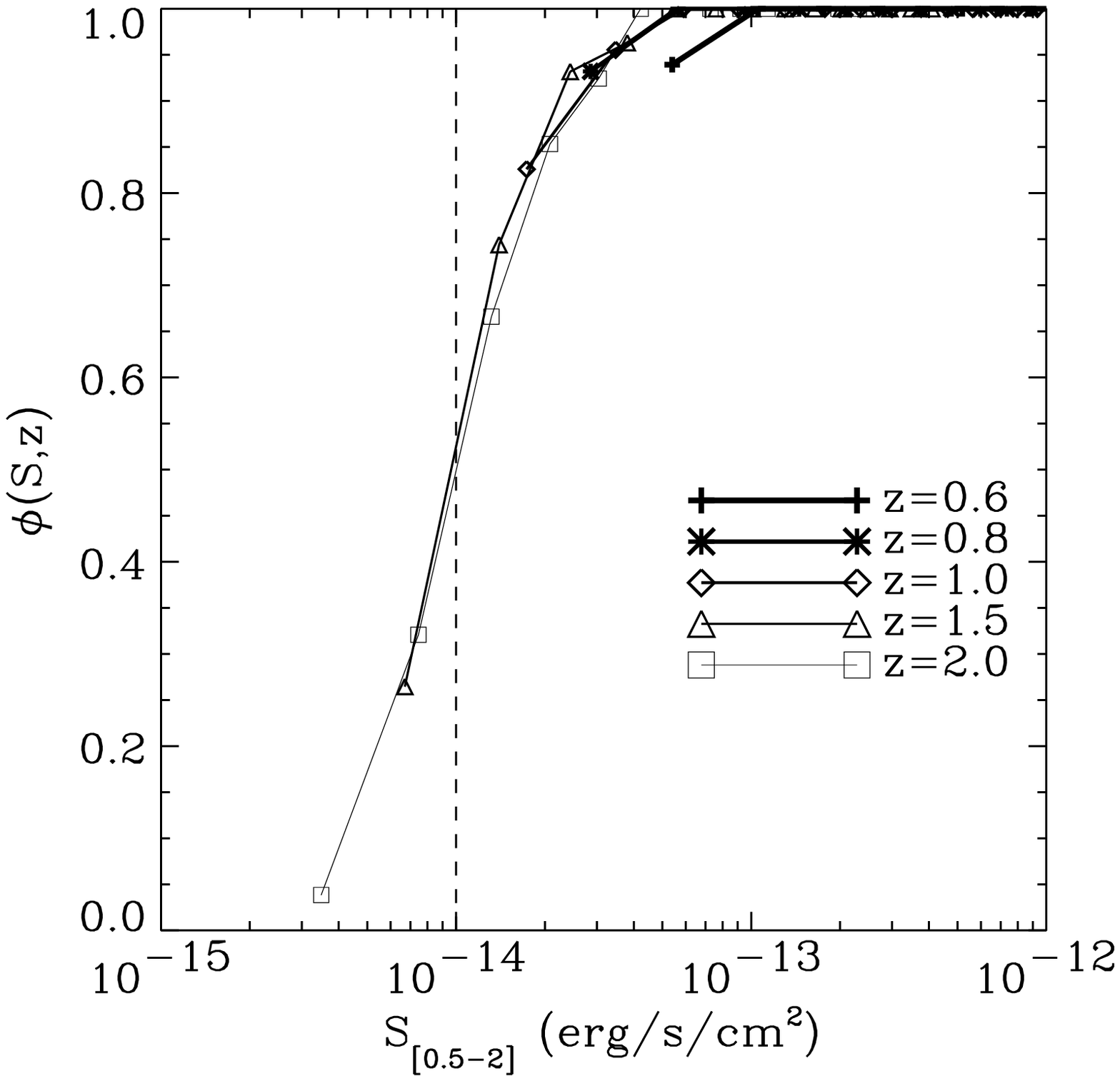,width=8cm}}
\caption{\label{fig:phi_sz}Selection function $\phi(S,z)$ expressed in
  terms of the flux $S$ in the [0.5-2] keV band for the \lcdm model.
  Only fluxes corresponding to temperatures above our 2 keV limit
  are displayed. The vertical line shows a flux limit of
  $10^{-14}$ erg s$^{-1}$ cm$^{-2}$; it illustrates deviations from a
  strict flux limit when realistic observing conditions are taken into
  account (esp. source confusion).}
\end{figure}

It is instructive to compare our selection function to that
corresponding to a constant flux limit, as assumed in many previous
studies. The flux $S_{[0.5-2]}$ of a cluster in the $[0.5-2]$ keV band
at redshift $z$ can be derived from its temperature using the $L-T$
relation (Eq.~[\ref{eq:l_t}]), the Raymond-Smith spectrum and the
luminosity-distance relation. Using this correspondence, we can
express the selection function in terms of the flux rather than
temperature. The resulting selection function $\phi(S_{[0.5-2]},z)$ is
shown in Fig.~\ref{fig:phi_sz}. Our selection function clearly
does not exactly correspond to a sharp flux limit, but is instead a
smooth function of the flux. For latter comparison, we note that
a detection probability of 50\% corresponds approximately to a flux
limit of $S_{[0.5-2]} = 10^{-14}$ \flux\ at the redshifts of
interest.

In the following, we will use our selection function which includes
most of the relevant instrumental and observational limitations.  For
comparison, we will also consider the selection function corresponding
to the above flux limit, along with that corresponding to temperature
limit $T>2$ keV. The latter selection is useful to study the effect of
the removal of small clusters on our predictions.

\section{Cluster Counts}
\label{counts} We first compute the expected cluster number counts
in the survey. This is done using the cluster selection function
derived in the previous section combined with the Press-Schechter
formalism. We first briefly review the main assumptions involved in
our calculation of the mass function and of the temperature function
of clusters. We then compute the expected projected number of clusters
on the sky, as function of redshift. Finally, we show how the
resulting redshift distribution constrains cosmological parameters
within CDM models.

\subsection{Mass Function}
\label{mass_function}
The Press-Schechter formalism provides an analytic expression for the
abundance of dark matter halos (Press \& Schechter \cite{ps74}).  At a
given redshift $z$, the differential number of dark matter halos of
mass $M$ per unit comoving volume is
\begin{equation}
\label{eq:dndm}
\frac{dn}{dM} = \sqrt{\frac{2}{\pi}} \frac{\overline{\rho}}{M}
\frac{d\nu}{dM} e^{-\frac{\nu^2}{2}}
\end{equation}
where $\overline{\rho}$ is the present mean matter density. The peak
height is defined as $\nu(M)=\delta_{c}/\sigma(M)$, where $\sigma(M)$
is the linear rms fluctuation in a sphere containing a mean mass $M$.
We compute $\sigma(M)$ for an arbitrary cosmological model by
integrating the linear power spectrum $P_{\rm lin}(k)$ derived from
the BBKS transfer function (Bardeen et al. \cite{bar86}; with the
conventions of Peacock \cite{pea97}), evolved with the linear growth
factor $D(z)$. The density threshold $\delta_{c}$ depends weakly on
cosmology (i.e. on $\Omega_{m}$ and $\Omega_{\Lambda}$) and was
computed using the fitting formulae of Kitiyama \& Suto
(\cite{kit96}).

\subsection{Temperature Function}
\label{t_function}
The X-ray temperature of a cluster at redshift $z$ is taken to be the
virial temperature which is given by (see e.g. Eke, Cole \& Frenk
\cite{eke96})
\begin{equation}
\label{eq:t_vir}
kT \simeq \frac{7.75}{\beta_{v}} (1+z) \Omega_{m}^{\frac{1}{3}}
\left( \frac{M}{M_{15}} \right)^{\frac{2}{3}}
\left( \frac{\mu}{0.59} \right)
\left( \frac{\Delta_{c}}{178} \right)^{\frac{1}{3}}
{\rm keV},
\end{equation}
where the average virial overdensity
$\Delta_{c}(z,\Omega_{m},\Omega_{\Lambda})$ can be evaluated using the
fitting formulae of Kitiyama \& Suto (\cite{kit96}), $M_{15} = 10^{15}
h^{-1} M_{\odot}$ and the value $\mu=0.59$ for the number of particles
per proton mass corresponds to a hydrogen mass fraction of 76\%. The
factor $\beta_{v}$ is equal to about 1 for a truncated singular
isothermal sphere. We adopt this value as it provides a good fit to
numerical simulations (Eke, Cole \& Frenk \cite{eke96}; Bryan \&
Norman \cite{bry97}).

Combining Eqs.~(\ref{eq:dndm}) and (\ref{eq:t_vir}) we can derive the
differential temperature function
\begin{equation}
\label{eq:dndt}
\frac{dn}{dT}=\frac{dn}{dM}\frac{dM}{dT}.
\end{equation}
It is often more convenient to consider the number density of clusters
with temperatures above a given minimum, $n(>T) = \int_{T}^{\infty}
dT'~\frac{dn(T')}{dT}$.

To illustrate the dependence of our prediction on cosmological
parameters, we consider the three cosmological models listed in
Tab.~\ref{tab:models}, i.e a tilted ($\tau$) CDM, \lcdm and
\ocdm. The normalization of these models is determined by $\sigma_{8}$,
the amplitude of mass fluctuations on $8 h^{-1}$ Mpc scale. Our chosen
numerical values correspond to the constraints derived from current
cluster surveys, namely $\sigma_{8} \simeq 0.52
\Omega_{m}^{-0.52+0.13\Omega_{m}}$ for the flat case and $\sigma_{8}
\simeq 0.52 \Omega_{m}^{-0.46+0.10\Omega_{m}}$ for the for the open
case (Eke et al. \cite{eke96}). The shape of the matter power spectrum
is controlled by the shape parameter $\Gamma$ which, unless otherwise
specified, we fix at 0.23, as indicated by galaxy clustering surveys
(see Viana \& Liddle \cite{via96} and reference therein).

\begin{table}
\caption{Cosmological Models}
\label{tab:models}
\begin{tabular}{llllll}
\hline
Model & h & $\Omega_{m}$ & $\Omega_{\Lambda}$ & $\sigma_{8}$ & $\Gamma$ \\
\hline
\tcdm    & 0.5 & 1   & 0   & 0.52 & 0.23 \\
\lcdm & 0.7 & 0.3 & 0.7 & 0.93 & 0.23 \\
\ocdm         & 0.7 & 0.3 & 0   & 0.87 & 0.23 \\
\hline
\end{tabular}
\end{table}


\subsection{Projected Cluster Counts}
From the temperature function $\frac{dn}{dT}$ (Eq.~[\ref{eq:dndt}]),
we can compute the projected surface density of clusters on the sky.
Noting that the comoving volume element is $dV=R^2d\chi d\Omega$,
where $d\Omega$ is the infinitesimal solid angle, $\chi$ is the
comoving radius and $R(\chi)$ is the comoving angular-diameter radius,
we find that the number of clusters per unit solid angle, temperature
and redshift interval is
\begin{equation}
\frac{dN}{dT dz} = R^{2} \frac{d\chi}{dz} \frac{dn}{dT},
\end{equation}
where $\frac{d\chi}{dz}=-\frac{c}{H_{0}} \left[ (1-\Omega) a^{-2}
  +\Omega_{m} a^{-3} +\Omega_{\Lambda} \right]^{-\frac{1}{2}}$ as
derived from the Friedmann equations. As a result the observed surface
density of clusters per unit redshift
\begin{equation}
\label{eq:dnobsdz}
\frac{dN_{\rm obs}}{dz} = \int dT  \frac{dN}{dT dz} \phi(T,z)
\end{equation}
where $\phi(T,z)$ is the survey selection function
(Eq.~[\ref{eq:phi}]).

\subsection{Predictions}
\begin{figure}
\centerline{
\psfig{figure=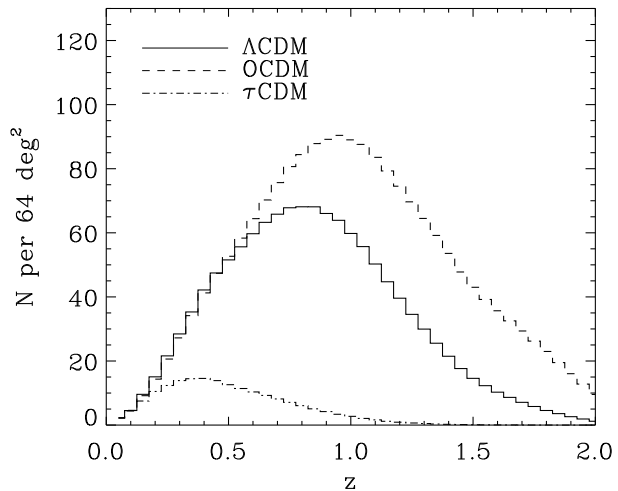,width=8cm}}
\caption{Projected number counts of clusters as a function of
  redshift in the three cosmological models. The selection function
  $\phi(T,z)$ for the XMM-LSS derived from image simulations was used
  for each model. The redshift bins have a width of $\Delta z=0.05$.
\label{fig:nz64}}
\end{figure}

The predicted projected counts as a function of redshift are shown on
Fig.~\ref{fig:nz64} for the three cosmological models whose parameters
are listed in Tab.~\ref{tab:models}. The counts correspond to the full
64 deg$^{2}$ of the completed XMM-LSS survey. The counts from the
three models agree at low redshifts ($z<0.2$), as expected since the
three models were normalized with the number of clusters in the local
universe. On the other hand, the number counts differ greatly at
larger redshifts. The \tcdm model predicts much smaller number of
clusters at $z>0.2$, while the \lcdm and \ocdm models differ for
$z>0.6$. The predicted counts at $z>1$ are larger in the \ocdm model
compared to that in the \lcdm case, due to the somewhat slower
evolution of the growth factor in the open model.  These differences
illustrate the well known fact that the evolution of cluster counts is
a powerful probe of $\Omega_{m}$ and $\Omega_{\Lambda}$. Note
that our predictions implicitly rely on the $L-T$ relation of
Eq.~(\ref{eq:l_t}), which was used to derive the selection function
$\phi(T,z)$ from the simulations (see \S\ref{simulations}).

In the 64 deg$^{2}$ of the survey, the expected number of detected
clusters with $0<z<1$ (and $T>2$ keV) is about 900, 1000 and 175
for the \lcdm, \ocdm and \tcdm model, respectively. For $1<z<2$, the
expected counts are about 400, 900 and $<10$, for each model
respectively.  Our total number of clusters for $0<z<2$ of about 1300
in the former model, is in approximate agreement with the predictions
of Moscardini et al. (\cite{mos00}) who predict about 1100 clusters in
the XMM-LSS for a \lcdm model with slightly different flux and
temperature thresholds.

Fig.~\ref{fig:n_sigma8} shows how these predictions for the \lcdm
model depend on $\sigma_{8}$ and $\Omega_{m}$.  Clearly, the number
counts are very sensitive on these two parameters.  Taking
$\sigma_{8}=0.93 \pm 0.07$ for $\Omega_m=0.3$ (Eke et al.
\cite{eke96}), the expected number of clusters in this model is
between 600-1200 for $0<z<1$ and 200-700 for $1<z<2$. 

\begin{figure}
\centerline{
\psfig{figure=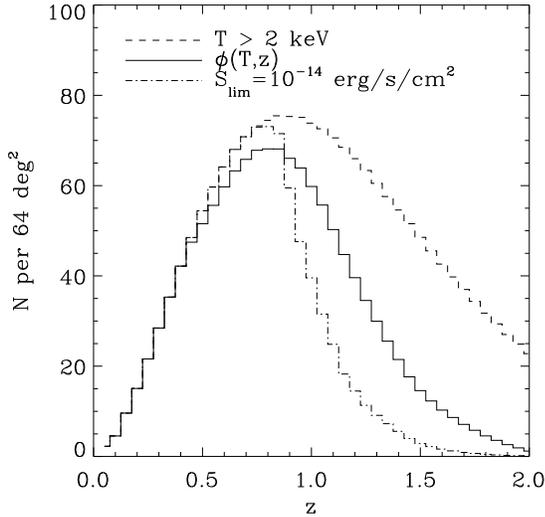,width=8.0cm}}
\caption{Projected number counts of clusters for the \lcdm model with
    different selection criteria: with the selection function $\phi(T,z)$
    (as in figure~\ref{fig:nz64}), with a temperature limit only ($kT>2$
    keV), and with a flux limit ($S_{[0.5-2]}> 10^{-14}$ erg
    s$^{-1}$ cm$^{-2}$). Again, the redshift bins have a width of $\Delta
    z=0.05$.
\label{fig:nz64_lcdm}}
\end{figure}

\begin{figure}
  \centerline{\psfig{figure=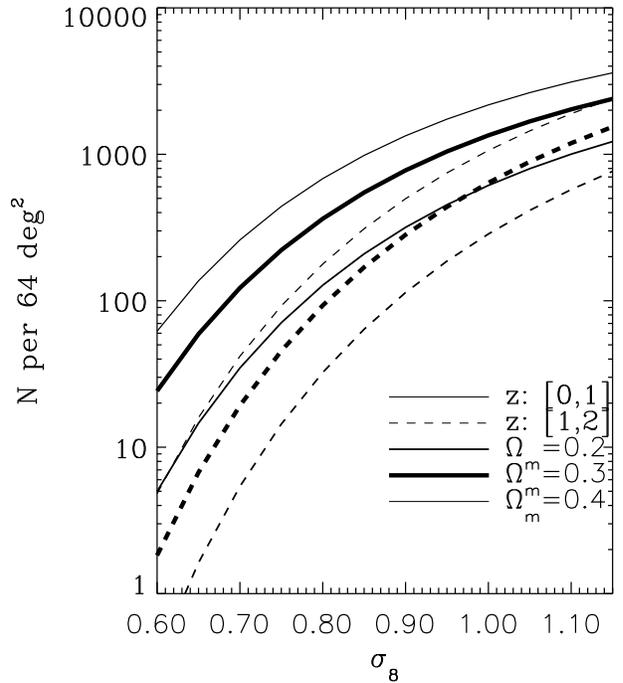,width=8cm}} \caption{Cluster
    counts expected for the XMM-LSS as a function of $\sigma_{8}$ and
    $\Omega_{m}$ in the \lcdm model. The XMM-LSS simulated selection
    function was used in all cases. Counts for the $0<z<1$ and $1<z<2$
    redshift intervals are shown as the solid and dashed lines,
    respectively. In each case, models with $\Omega_{m}=0.4$, 0.3 and
    0.2, are shown from top to bottom, respectively. 
\label{fig:n_sigma8}}
\end{figure}

The effect of the selection function on these predictions are shown in
Fig.~\ref{fig:nz64_lcdm}. The number counts for the \lcdm model are
shown as a function of redshift, for our selection function
$\phi(T,z)$ (as in Fig.~\ref{fig:nz64}), for the flux limited case and
for the temperature limited case. For $z \ga 0.9$, the counts with the
selection function are larger than the flux limited counts. This is
expected, since a large number of clusters fainter than the flux limit
contribute to the $\phi(T,z)$ counts because of the tail of the
selection function (see Fig.~\ref{fig:phi_sz}). This demonstrates the
importance of considering all the observational details when making
such predictions, especially confusion by point-like sources, an
unavoidable drawback for a highly sensitive instrument. Notice that
the temperature limited counts agree with the other two for $z<0.4$.
This shows that our counts are limited mainly by the temperature at
low redshifts. Not surprisingly, the $T$-limited counts overpredict
the abundances for $z>0.4$. As a check, we have compared our
predictions with that of Romer et al. (\cite{rom01}). For their
effective flux limit of $S_{[0.5-2]}>1.5\times 10^{-14}$ erg s$^{-1}$
cm$^{-2}$ and $kT>2$ keV, we find that our predictions agree very well
with the counts shown in their Figure~6c.

Note that the fluxes of the clusters measured in XMM-LSS will be
rather uncertain (see VPG). This will prevent us from making accurate
determinations of the luminosity, and therefore of the temperature and
of the mass, of each cluster. In this paper, we therefore
consider statistical quantities (such as the counts and the
correlation function) which are averaged over the population of
clusters detectable in the survey, and therefore do not require this
information. A study of the further constraints which can be derived
from the (uncertain) flux measurements is left to future work.

\subsection{Cosmological Constraints from Cluster Counts}
\label{nz:constraints}

We now wish to study how the cluster counts can be used to constrain
cosmological parameters. For this purpose, we generated cluster counts
from the predicted counts from Eq.~(\ref{eq:dnobsdz}) in \lcdm model
for several redshift bins (as shown in Fig.~\ref{fig:nz64} for
different models). We used the XLSS selection function $\phi(T,z)$ and
survey area, and considered redshift bins of width $\Delta z=0.1$ from
$z=0$ to $z=1$ assuming Poissonian errors for the number counts in
each bin. Note that the Poisson errors neglect the uncertainty arising
from cosmic variance; this is a good approximation since this latter
uncertainty is small for a survey area as large as that of the XLSS
(compare the correlation length $\sim 20 h^{-1}$ Mpc to the survey
size, $\sim 200 h^{-1}$ Mpc at $z \simeq 0.5$). We then calculated the
$\chi^{2}$-difference between the simulated counts and that expected
in a \lcdm model for a range of values for $\sigma_{8}$ and
$\Omega_{m}$. This approach is rather computer intensive, but it is
more accurate than the faster Fisher matrix method, which approximates
the confidence regions as ellipses and can be incorrect when the
parameter space is non-Gaussian (Holder et al.  \cite{hol01}). 
Note that our $\chi^{2}$ fitting method is only strictly valid for
gaussian errors. This is however a good approximation in our
case, given the large number of clusters in the redshift bins shown in
Figure~\ref{fig:nz64}, further enhanced by our use of $\Delta z=0.1$
bins rather than $0.05$ in the Figure.

\begin{figure}
  \centerline{ \psfig{figure=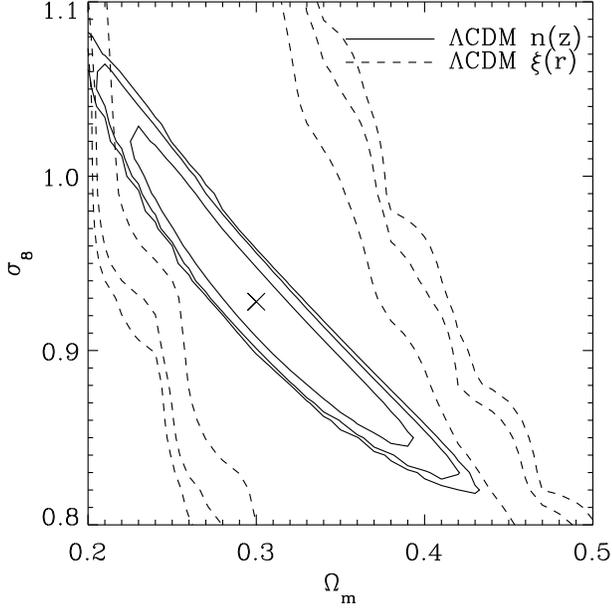,width=8.0cm}}
    \caption{Constraints on the cosmological parameters $\Omega_m$ and
      $\sigma_8$ for the \lcdm model, from the cluster counts
      (solid lines) and from the correlation function (dashed lines).
      In each case, the 68\%, 90\% and 95\% confidence level contours are
      shown along with the assumed model (cross).
      \label{fig:om_s8_lcdm}}
\end{figure}

The resulting constraints on these cosmological parameters are shown
as the solid lines in Fig.~\ref{fig:om_s8_lcdm}. The input \lcdm model
is that of Tab.~\ref{tab:models} and is shown with a cross. The 90\%
and 95\% confidence regions are shown as solid lines. The cluster
counts alone will provide tight constraints on both parameters, with
95\% uncertainties of about 0.06 and 0.05 for $\sigma_{8}$ and
$\Omega_{m}$, respectively. As can be seen from the elongation of the
contours, the two parameters are however somewhat degenerate.  Any
additional information on either parameters, can thus be used to
reduce their respective uncertainty. Note that these constraints
only reflect statistical errors, and neglect potentially important
systematic uncertainties in the scaling laws of clusters. A discussion
of the limitations imposed by systematics on our predictions is
presented in \S\ref{systematics}.  


\begin{figure}
  \centerline{ \psfig{figure=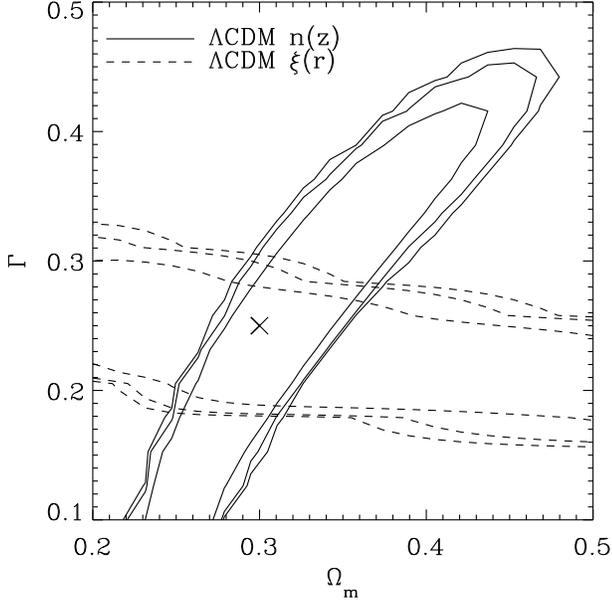,width=8cm}}
    \caption{Constraints on $\Gamma$ and $\Omega_m$ for the \lcdm,
      from the cluster counts (solid lines) and from the
      correlation function (dashed lines). As before, the 68\%, 90\%
      and 95\% confidence level contours are shown (solid lines) along
      with the assumed model (cross).
      \label{fig:gamma_omega}}
\end{figure}

It is also interesting to investigate how cluster counts can constrain
$\Gamma$, the shape parameter of the matter power spectrum.  For this
purpose, we follow the same $\chi^2$ procedure, this time varying
$\Gamma$ and $\Omega_{m}$, while $\sigma_{8}$ follows the Eke et
al. (\cite{eke96}) relation (see \S\ref{t_function}). The resulting
confidence contours for the \lcdm model are shown as the solid lines
in Fig.~\ref{fig:gamma_omega}. Clearly, these two parameters are quite
degenerate with counts alone, hampering the determination of
$\Omega_{m}$, which can only be determined with an accuracy of about
40\% (95\% CL).  More information is therefore required to alleviate
these limitations.  One obvious possibility is to use other measures
of large-scale structure such as galaxy catalogues to constrain
$\Gamma$.  This has the disadvantage of relying on assumptions about
the bias of galaxies and on an external data set. In the next section,
we show how the degeneracy can be broken by measuring the correlation
function of the galaxy cluster population.

\section{Cluster Correlation function}
\label{correlation}
We now turn to the cluster correlation function which quantifies the
spatial clustering of clusters in the survey. We first use the
extended halo model of Mo \& White (\cite{mo96a}) to predict the
observed cluster correlation function at a given redshift. We then
consider the average correlation function in a finite redshift range,
compute its uncertainties, and study the constraints its measurement
can place on cosmological parameters.

\subsection{Observed Correlation Function : at a single redshift}
According to the Mo \& White (\cite{mo96a}) halo formalism, the
correlation function of two sets of clusters with masses $M$ and $M'$
and with comoving separation $r$ is given by
\begin{equation}
\xi(r,M,M',z) \simeq b(M,z) b(M',z) \xi_{\rm lin}(r,z),
\end{equation}
where $\xi_{\rm lin}(r,z)=\xi_{\rm lin}(r,0)D^{2}(z)$ is the linear
correlation function which is the Fourier Transform of the linear
power spectrum $P_{\rm lin}(k,z)$. The mass dependent bias parameter
of the halos is given $b(M,z)=1+(\nu^2-1)/\delta_c$, with the
conventions of \S\ref{mass_function}. Note that the separation of the
clusters is assumed to be small compared to the scale in which any
evolution takes place.

It is easy to show that the resulting observed correlation function in
a narrow redshift interval is given by (Suto et al. \cite{sut00};
Moscardini et al. \cite{mos00} and reference therein)
\begin{equation}
\xi_{\rm obs}(r,z)=b_{\rm eff}^{2}(z) \xi_{\rm lin}(r,z),
\end{equation}
where the effective bias is
\begin{equation}
b_{\rm eff}(z)= \int dM \frac{dn_{\rm obs}}{dM} b(M) \left/
\int dM \frac{dn_{\rm obs}}{dM} \right. ,
\end{equation}
and where the observed differential number counts are given by
$\frac{dn_{\rm obs}}{dM}=\frac{dn}{dT}\frac{dT}{dM}\phi(T,z)$. Note
that, in our analysis, we neglect redshift-space distortions which
were shown to yield only about 10\% corrections on the amplitude of
the correlation function (Suto et al. \cite{sut00}; Moscardini et al.
\cite{mos00})

In general, the evolution of $\xi_{\rm obs}(r,z)$ is determined by two
competing effects. First, the growth of structures induces the linear
correlation function $\xi_{\rm lin}(r,z)$ to decrease as the redshift
increases. On the other hand, the clusters which are detectable at large
redshifts are more massive and therefore more strongly biased. The
effective bias $b_{\rm eff}(z)$ thus tends to be larger at high
redshift. To study the interplay between these two effects it is
convenient to define
\begin{equation}
b_{\rm eff,0}(z) \equiv \left[ \frac{\xi_{\rm obs}(r,z)}
{\xi_{\rm lin}(r,0)}\right]^{\frac{1}{2}}
= b_{\rm eff}(z) D(z).
\end{equation}
This quantity provides the bias of the observed cluster correlation
with respect to the linear correlation function at $z=0$, and
therefore quantifies the evolution of the correlation function.

\begin{figure}
  \centerline{ \psfig{figure=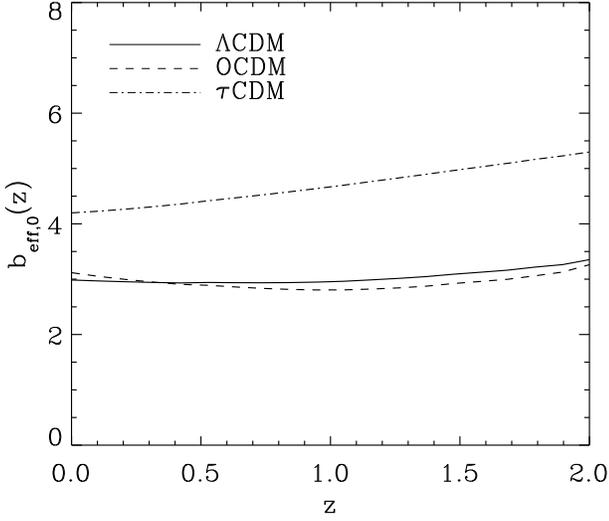,width=8cm}} \caption{Evolution
   of $b_{\rm eff,0}(z)$, the effective bias of the cluster
   correlation function $\xi_{\rm obs}(r,z)$ with respect to the
   linear mass correlation function $\xi_{\rm lin}(r,0)$ at $z=0$.
   This quantifies the evolution of the cluster correlation function.
   It is shown for the three cosmological models as a function of
   redshift.  \label{fig:beff0} }
\end{figure}

\begin{figure}
  \centerline{ \psfig{figure=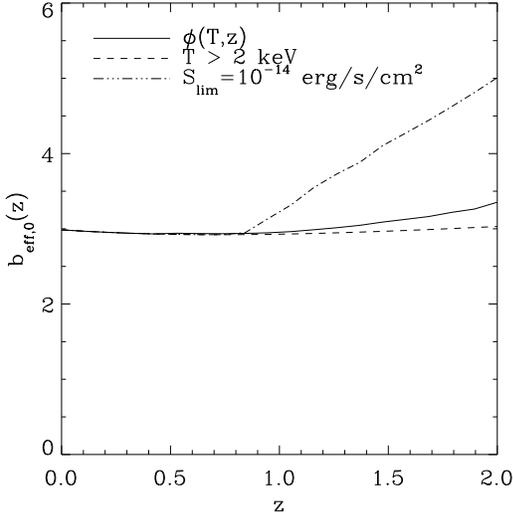,width=8cm}}
  \caption{Evolution of $b_{\rm eff,0}(z)$ in the \lcdm
    model for different selection
    schemes: the XMM-LSS selection function, a
    temperature limited survey, and a flux limited survey.
    \label{fig:beff0_lcdm} }
\end{figure}

The behaviour of $b_{\rm eff,0}(z)$ is shown on Fig.~\ref{fig:beff0}
for the three cosmological models and for the XMM-LSS selection
function. The curves are remarkably flat for all models, showing that
the cluster correlation function evolves only very weakly from $z=0$
to $z=2$. It is interesting to study whether this lack of evolution
depends on the selection function.  Fig.~\ref{fig:beff0_lcdm} shows
$b_{\rm eff,0}(z)$ for each selection scheme for the \lcdm model. The
evolution is also very weak for the temperature limited case. For
the flux-limited sample, the evolution is stronger for $z>1$. This
evolution is lost when the full selection function $\phi(T,z)$ is
used. In all cases, there is effectively no evolution for $0<z<1$.  We
shall thus, in the following, derive the constraints on cosmology
integrating the correlation function over the $0<z<1$ range.

\subsection{Observed Correlation Function: Redshift Average}
To maximize the sensitivity, it is useful to measure the correlation
function over a wide redshift range. In this case, we must take into
account the pair-weighted evolution of the correlation function
$\overline{\xi}(r,z)$ within the light-cone section of interest.  The
average correlation function in the redshift interval $z_{\rm
  min}<z<z_{\rm max}$ is thus given by (Suto et al. \cite{sut00};
Moscardini et al. \cite{mos00} and reference therein)
\begin{equation}
\overline{\xi}_{\rm obs}(r) = \frac{
\int_{z_{\rm min}}^{z_{\rm max}}
  dz~\frac{d\chi}{dz} R^{2} n_{\rm obs}^{2}(z)
  \xi_{\rm obs}(r,z)}
{\int_{z_{\rm min}}^{z_{\rm max}} dz~\frac{d\chi}{dz} R^{2}
  n_{\rm obs}^{2}(z)},
\end{equation}
where the observed number density of clusters is $n_{\rm obs}(z) =
\int~dT \frac{dn}{dT} \phi(T,z)$.

\begin{figure}
  \centerline{ \psfig{figure=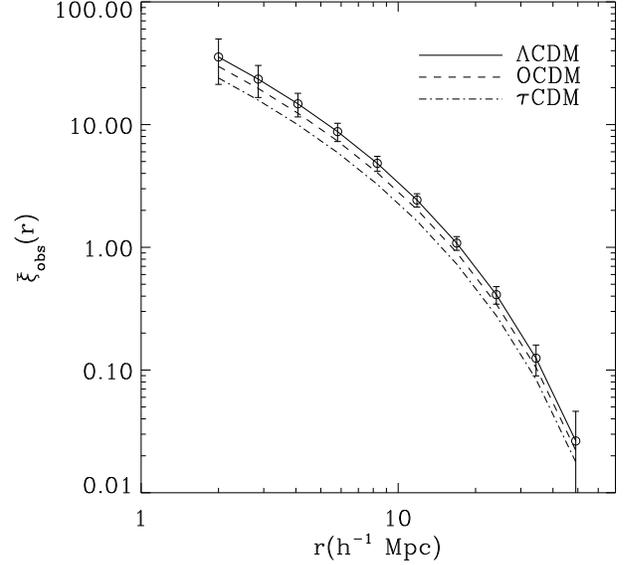,width=8cm}}
    \caption{Prediction for the cluster correlation function averaged
      in redshift interval $0<z<1$. The three cosmological models are
      shown.  For clarity, the expected measurement error bars for the
      full 64 deg$^{2}$ of the XMM-LSS are shown only for the
      \lcdm model.
    \label{fig:xiobs}. }
\end{figure}

The resulting correlation functions for the three cosmological models
are shown on Fig.~\ref{fig:xiobs}. The XMM-LSS selection function
was used in all cases, along with a redshift range of $0<z<1$. The
correlation functions for the 3 models have very similar shape, as
expected since the same value for $\Gamma$ was assumed in all cases.
Notice that the \tcdm correlation function has a lower amplitude
than that for the other two models.  This is expected since the former
model has a lower value for $\sigma_{8}$ (see Tab.~\ref{tab:models}).
The \lcdm and \ocdm models have very similar values of
$\sigma_{8}$ and thus yield correlation functions with very similar
amplitudes. 

Our predictions are in good agreement with the results of
Moscardini et al. \cite{mos00}. In particular, for our \lcdm model,
these authors find a correlation length $r_{0}$ (defined by
$\overline{\xi}_{\rm obs}(r_{0})\equiv 1$) of about 13 and 15 h$^{-1}$
Mpc for $z<0.3$ and $z>0.3$, respectively. This is to be compared with
our value of $r_{0} \simeq 17$ h$^{-1}$ Mpc for $z<1$.  We also
verified that the evolution of the correlation function is weak if we
adopt their assumed strict flux limit of $S_{[0.5-2]} = 5 \times
10^{-15}$ \flux\ .

\subsection{Uncertainties}
We now wish to estimate the uncertainty involved in measuring the
cluster correlation function. The simplest way to measure the
correlation function is to count the number of pairs in excess of
random in the survey volume. (This is sometimes taken as the
definition of the correlation function). An estimator for
$\overline{\xi}_{\rm obs}(r)$ in a bin of radii between $r$ and
$r+\Delta r$ is thus
\begin{equation}
\hat{\overline{\xi}}_{\rm obs}(r,\Delta r) \equiv
\frac{N_{\rm pairs}^{\rm obs}(r,\Delta r)}
{N_{\rm pairs}^{\rm ran}(r,\Delta r)} - 1,
\end{equation}
where $N_{\rm pairs}^{\rm obs}$ and $N_{\rm pairs}^{\rm ran}$ are the
number of pairs (i.e. with clustering) and for a random distribution
(i.e. without clustering), respectively, in the observed radius
interval. Ignoring boundary effects, the random number of pairs is
given by
\begin{equation}
N_{\rm pairs}^{\rm ran}(r,\Delta r) \simeq
2\pi A r^{2} \Delta r \int_{z_{\rm min}}^{z_{\rm max}}
dz \frac{d\chi}{dz} R^{2} n_{\rm obs}^{2}(z),
\end{equation}
where $A$ is the solid angle of the survey.

For weak signals, the error in measuring $\hat{\overline{\xi}}_{\rm
  obs}(r,\Delta r)$ will be dominated by Poisson statistics, and will
thus be given by (see e.g. Peebles \cite{pee80})
\begin{equation}
\label{eq:xihat_error}
\sigma[\hat{\overline{\xi}}_{\rm obs}(r,\Delta r)]
\simeq \frac{1 + \xi(r)}{\sqrt{ N_{\rm pairs}^{\rm obs}(r,\Delta r)}}
= \sqrt{ \frac{1 + \xi(r)}{N_{\rm pairs}^{\rm ran}(r,\Delta r)}}.
\end{equation}
This provides us with an expression for the error in the correlation
function for a finite survey. The resulting errors for the \lcdm model
and for the full 64 deg$^{2}$ of the XMM-LSS are shown in
Fig.~\ref{fig:xiobs}. Again, these errors only reflect
statistical errors and neglect uncertainties in the cluster scaling
relations. A discussion of the impact of systematics, including that
arising from spatial variations of the survey sensitivity, is
presented in \S\ref{systematics}.


\subsection{Combined Cosmological Constraints}
\label{cosmo_counts} 
We now study how the measurement of the cluster correlation function
constrains cosmological parameters. As for the cluster counts
(\S\ref{cosmo_counts}), we use a $\chi^{2}$-fit to simulated
measurements of the correlation function to derive confidence regions
in parameter space. We considered a redshift interval of $0<z<1$ and
computed the errors using Eq.~(\ref{eq:xihat_error}).

The resulting confidence contours for a joint fit of $\Omega_{m}$ and
$\sigma_{8}$ are shown as the dashed lines in
Fig.~\ref{fig:om_s8_lcdm}, for the \lcdm model. The constraints from
the correlation function are rather weak on this plane alone, with
little dependence on $\sigma_{8}$.  These constraints are however
somewhat orthogonal to that from cluster counts and are thus
complementary.

The constraints from the correlation function on $\Gamma$ and
$\Omega_{m}$ are shown in Figure~\ref{fig:gamma_omega}. The shape
of the resulting confidence contours agrees qualitatively with the
predictions of Moscardini et al. (\cite{mos00}) for the shallower
REFLEX survey. The contours show little dependence on
$\Omega_{m}$ and are therefore nearly orthogonal to that from cluster
counts alone.  With the combined counts and correlation function
constraints, $\Omega_{m}$, $\sigma_{8}$ and $\Gamma$ can be measured
with a precision of about 15\%, 10\%, and 35\%, respectively at the
95\% confidence level. The cluster correlation function thus provides
the required information to break the degeneracy present when cluster
counts alone are considered.  The impact of systematics on these
constraints is discussed in \S\ref{systematics}. 

\begin{figure}
  \centerline{ \psfig{figure=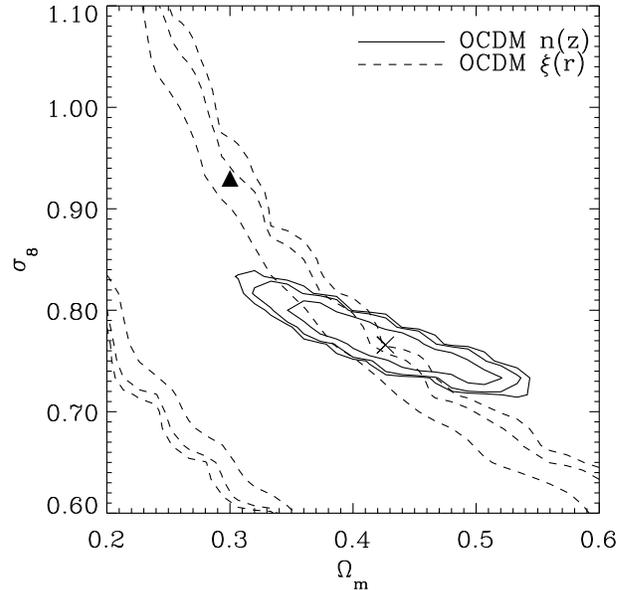,width=8.0cm}}
    \caption{Constraints on $\Omega_m$ and $\sigma_8$ resulting
      from a fit of the \ocdm model to the \lcdm predictions. Both the
      cluster counts (solid lines) and the correlation function
      (dashed lines) are shown. In each case, the 68\%, 90\% and 95\%
      confidence level contours are shown, along with the $\chi^2$
      minimum (cross) and the input values for the \lcdm model (filled
      triangle).
    \label{fig:om_s8_ocdm} }
\end{figure}

Until now, we have focused on the \lcdm model, thus implicitely
assuming prior knowledge that the universe is flat. While flatness is
strongly indicated by measurements of the Cosmic Microwave Background
anisotropies (e.g. Jaffe et al.~\cite{jaf01}), it is interesting to
establish whether our predictions depend on this assumption. To study
this, we fitted an \ocdm model to the \lcdm predictions for both the
cluster counts and the correlation function. The resulting constraints
on $\sigma_{8}$ and $\Omega_{m}$ are shown in
Fig.~\ref{fig:om_s8_ocdm}. For the cluster counts, the best fit values
for both of these parameters are now biased (compare the best fit,
cross, to the input values, triangle). This bias again hampers the
determination of $\Omega_{m}$ using cluster counts alone. Thankfully,
the correlation function can again help overcome this limitation.
Indeed, the contours for $\xi(r)$ are now somewhat incompatible with
that from the counts (compare to Fig.~\ref{fig:om_s8_lcdm}). The
consistency between the counts and correlation function constraints
can thus be used as a diagnostic and as a discriminant between the
\lcdm and the \ocdm model.

\section{Systematic Uncertainties}
\label{systematics}
So far in our analysis, we have neglected systematic
uncertainties. Here we review the major source of systematics and
discuss their impact on our predictions.\footnote{We thank Kathy
Romer, the referee, for her many suggestions which are included in this
section.}

First, we have assumed that the $L-T$ and $M-T$ relations were
perfectly known. Physical processes such as feedback from star
formation and reionization, can modify these relationships and
therefore yield detectable changes in the cluster counts and
correlation function (see e.g. Muanwong et al. \cite{mua01} and
reference therein). These could perhaps explain the discrepancy which
has recently been found between the observed $M-T$ relation and that
derived from numerical simulations (see Finoguenov, Reiprich, \&
B\"{o}hringer \cite{fin01}; Allen, Schmidt, \& Fabian \cite{all01} and
reference therein). Cluster mergers have also been shown to produce
sharp temperature and luminosity jumps, thus likely affecting the
$L-T$ relation (Ricker \& Sarazin \cite{ric01}). Another important
issue is the existence of high redshift cooling flows which may affect
the detectability of distant clusters and their $L-T$ relation
(e.g. Henry \cite{hen00}). 

By the time the XMM-LSS is completed, better insights will be gained
about these various physical effects, thanks to deep pointed
observations of clusters with XMM and Chandra. An interesting way of
parameterizing the residual uncertainties has been proposed by Diego
et al. (\cite{dieg01}). These authors kept the normalisation and
evolution of the different cluster scaling relations as free
parameters, and studied the resulting degeneracies with cosmological
parameters. It would be interesting to extend their approach to
include constraints from the cluster correlation function. The study
of these pending theoretical questions is left for future work.

In our simulation, we have also assumed that the core radius $r_{c}$
is fixed. More realistically, the core radius could vary with mass and
redshift. For instance, Jones et al. (\cite{jon99}) found a
correlation between core radii and temperatures in a cluster sample
from the {\it Einstein} mission. Our detection sensitivity depends
most strongly on the cluster flux, which is independent of the core
radius in our simulations. A varying $r_{c}$ is therefore unlikely to
affect our selection function very much, except perhaps in our ability
to separate clusters from point sources.

Another systematic effect arises from our assumption of a sharp
temperature cutoff ($kT>2$ keV) for our cluster selection.  Since
temperatures will not be available with XMM-LSS, this will be enforced
in practice by setting a luminosity cutoff, derived from the observed
fluxes and redshifts. Because the luminosity function is very steep, the
errors in the observed fluxes will lead to a bias tending to include
low mass clusters in the counts. This effect can be corrected for by
running simulations similar to ours, but which would include objects
with $kT<2$ keV.

\begin{figure}
\centerline{
\psfig{figure=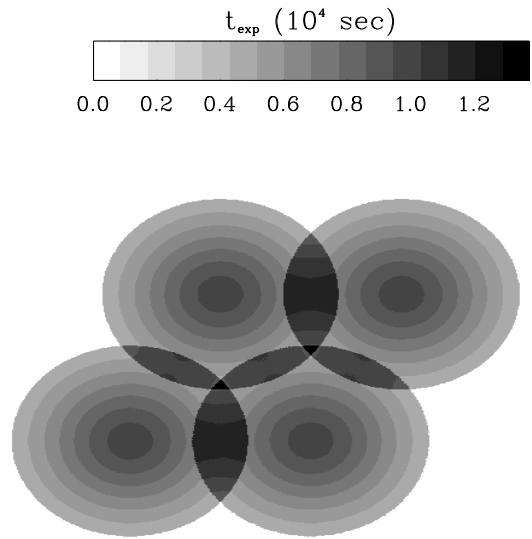,width=8cm}}
\caption{\label{fig:texp}Exposure time map for the XMM-LSS tiling
pattern. Four XMM fields are displayed out to a maximum radius of 13'.
The fields are separated by 20 arcmins from each other.}
\end{figure}

Another complication is induced by variations in the detection
sensitivity over the survey area. Figure~\ref{fig:texp} shows the
effective exposure time for the XMM-LSS tiling strategy, in which the
10 ksec exposures are separated by 20 arcmin. As is apparent in the
figure, vignetting causes variations in the exposure time of about
50\%, corresponding to flux sensitivity variations of about 25\%. This
will affect both the uniformity of the cluster counts and the
observed cluster correlation function. Since the vignetting function
is well calibrated, the sensitivity variations can however be
accurately predicted and corrected for. This can be done by using
further image simulations to compute the selection function as
a function of position. Note that, in the simulations we have presented
in \S\ref{simulations}, we have only computed the average selection
function by placing clusters throughout the inner 13' of a single
XMM-EPIC pointing. Our detection limits are thus conservative, as they
do not include the increased sensitivity afforded in the regions of
overlap of two XMM fields.

\section{Conclusions}
\label{conclusion} 
Following the REFLEX survey (Guzzo et al. \cite{guz99}, B\"ohringer et
al. \cite{boh01}) based on the ROSAT All-Sky-Survey and the NEP ROSAT
survey (Henry et al. \cite{hen01}), the XMM-LSS survey will be about
1000 and 10 times deeper, respectively, and thus opens wide prospects
for cosmology. Indeed, it will provide an independent measurement of
cosmological parameter and thus complements Cosmic Microwave
Background and Supernova experiments. It will also provide a test of
important ingredients of the standard cosmological model, such as the
gravitational instability paradigm and the gaussianity of initial
fluctuations.

To study the constraints XMM-LSS will set on cosmological parameters,
we first derived the selection function for the survey using detailed
simulations of cluster detection in XMM-Newton images. We found that
our selection function differs significantly from a simple flux-limit
selection. We then computed the expected number counts of clusters in
several CDM models. We found that, for the currently favoured \lcdm
model, we expect about 600-1200 clusters to be detectable in XMM-LSS
at $0<z<1$ and about 200-700 at $1<z<2$, the uncertainty being
dominated by the current errors on $\sigma_{8}$ and $\Omega_{m}$.

Clusters counts beyond $z>0.5$ depend on cosmological parameters and
can thus be used to constrain models. Within a \lcdm model, we found
that the redshift dependence of the XMM-LSS cluster counts will allow
us to measure $\sigma_{8}$ and $\Omega_{m}$ with a precision of about
6\% and 18\% (95\% CL), respectively, if the shape parameter $\Gamma$
is known. In the absence of prior knowledge on $\Gamma$, the precision
on these parameters degrades considerably, if only cluster counts are
considered.


This limitation can be circumvented by considering the cluster
correlation function. One of the strength of XMM-LSS is indeed to
offer a uniform coverage over a wide contiguous area (64 deg$^{2}$),
with an extensive spectroscopic follow-up. This will allow us to
measure the correlation function in several redshift bins out to
$z=1$. Using the extended halo formalism of Mo \& White
(\cite{mo96a}), we computed the correlation function of clusters
detectable in XMM-LSS. We find that, for the selection function of
XMM-LSS, the correlation function is not expected to evolve
significantly from $z=0$ to 2. This results from the competing effects
of the growth of mass perturbations and the stronger bias of the
detectable massive clusters at large redshifts. It will thus be
important to compare the correlation function measured in two redshift
bins between $z=0$ and $z=1$: the verification of the lack of
evolution provides a test of the bias model for haloes and of the
gravitational instability paradigm.

The amplitude and shape of the cluster correlation function can be
used to lift the degeneracies present when cluster counts alone are
considered.  Within a \lcdm model, the correlation function function
measured in XMM-LSS at $0<z<1$, combined with the cluster counts, will
constrain $\Omega_{m}$, $\sigma_{8}$ and $\Gamma$ with a precision of
about 15\%, 10\% and 35\%, respectively (95\% CL).  Moreover, the
combination of the counts and the correlation function will provide a
consistency check for the \lcdm model, and a discrimation between this
model and the \ocdm model. The XMM-LSS therefore has great
potentials for the measurement of cosmological parameters. Note
however that the above constraints only reflect statistical errors and
ignore the limitations arising from systematic
uncertainties. Instrumental systematics, such as sensitivity
variations, could be important but can be accurately corrected using
further image simulations. Systematic uncertainties in the cluster
scaling relations may also contribute significantly to the error
budget. By the time the XMM-LSS is completed, better knowledge of
these scaling relations will have been derived from deeper pointed
observations with XMM and Chandra. The study of the impact of potential
residual uncertainties on cosmological constraints with
XMM-LSS is left for future work.

\section*{Acknowledgements}
We thank Kathy Romer, the referee, for her useful comments and
criticisms, and for her numerous suggestions regarding systematic
effects. AR was supported by a TMR postdoctoral fellowship from the
EEC Lensing Network, and by a Wolfson College Research Fellowship.


\begin{thebibliography}{99}
\bibitem[2000]{ada00}
  Adami, C., Ulmer, M.P., Romer, A.K., Nichol, R.C., Holden, B.P., 
  Pildis, R.A., 2000, ApJS, 131, 391 
\bibitem[2001]{all01}
  Allen, S.W., Schmidt, R.W., \& Fabian, A.C., 2001, MNRAS, 328, 37
\bibitem[1996]{xspec}
  Arnaud, K.A., 1996,  in ASP Conf. Ser., Vol. 101, Astronomical Data
  Analysis Software and Systems V,eds. Jacoby G.H. \& Barnes J. (San
  Francisco: ASP), 17 (XSPEC)
\bibitem[1999]{arn99}
  Arnaud, M. \& Evrard, A., 1999, MNRAS, 305, 631
\bibitem[1986]{bar86}
  Bardeen, J.M., Bond, J.R., Kaiser, N., \& Szalay, A.S., 1986, ApJ,
  304, 15
\bibitem[1996]{sex}
  Bertin, E., Arnouts, S., 1996, A\&AS, 117, 393
\bibitem[2001]{bor01b} 
  Borgani, S., \& Guzzo, L. 2001, Nature, 409, 39
\bibitem[2001]{boh01} 
  B\"ohringer, H., Schueker, P., Guzzo, L. et al., 2001, A\&A, 369,
  826 
\bibitem[1997]{bry97}
  Bryan, G.L., \& Norman, L., 1997, ASP Conf Ser. 123:
  Computational Astrophysics; 12th Kingston Meeting on Theoretical
  Astrophysics, 363
\bibitem[1976]{cff76}
  Cavaliere, A., Fusco-Femiano, R., 1976, A\&A, 49, 137
\bibitem[2000]{coll00}
  Collins, C. A., Guzzo, L., B\"ohringer, H.,
  Schuecker, P., Chincarini, G.,
  Cruddace, R., De Grandi, S.,
  MacGillivray, H. T., Neumann, D. M.,
  Schindler  S., Shaver  P., Voges  W., 2000, MNRAS, 319, 939
\bibitem[2001]{dieg01}
  Diego,  J.M.,  Mart\'inez-Gonz\'ales,  E.,  Sanz,  J.L.,
  Cay\'on, L., \& Silk, J., 2001, MNRAS, 325, 1533
\bibitem[1996]{eke96}
  Eke,  V.R., Cole, S., \& Frenk, C.S., 1996,  MNRAS, 282, 263
\bibitem[1998]{eke98}
  Eke, V.R., Navarro, J.F., \& Frenk, C.S., 1998, ApJ, 503, 569
\bibitem[2001]{fin01}
  Finoguenov, A., Reiprich, T.H., \& B\"{o}hringer, H., A\&A, 368, 749
\bibitem[2001]{gia01}
  Giacconi, R., Rosati, P., Tozzi, P. et al., 2001, ApJ, 551, 624
\bibitem[1999]{guz99}
  Guzzo, L., B\"ohringer, H., Schuecker, P. et al., 1999, ESO
  Messenger, 95, 27  
\bibitem[2001]{hai01}
  Haiman, Z., Mohr, J.J., \& Holder, G., 2001, ApJ, 553, 545
\bibitem[2001]{has01}
  Hasinger, G., Altieri, B., Arnaud, M., et al. 2001, A\&A, 365, L45
\bibitem[2000]{hen00}
  Henry, J.P., 2000, ApJ, 534, 565
\bibitem[2001]{hen01}
  Henry, J.P., et al. 2001, ApJ, 553, L109
\bibitem[2001]{hol01}
  Holder, G., Haiman, Z., Mohr, J.J., 2001, ApJL, submitted
  (astro-ph/0105396)
\bibitem[2001]{jaf01} Jaffe, A., et al. 2001, Phys.Rev.Lett. 86, 3475
\bibitem[1999]{jon99}
  Jones, C., \& Forman, W., 1999, ApJ, 511, 65
\bibitem[1996]{kit96}
  Kitiyama, T., \& Suto, Y., 1996, ApJ, 469, 480
\bibitem[1997]{kit97}
  Kitayama, T., \& Suto, Y., 1997, ApJ, 490, 557
\bibitem[1996]{mo96a}
  Mo, H.J. \& White, S.D.M., 1996, MNRAS, 282, 347
\bibitem[1996]{mo96b}
  Mo, H.J., Jing, Y.P. \& White, S.D.M., 1996, MNRAS, 282, 1096
\bibitem[2000]{mos00}
  Moscardini, L., Matarrese, S., \& Mo, H.J., 2000, astro-ph/0009006
\bibitem[2001]{mua01}
  Muanwong, O., Thomas, P.A., Kay, S.T., Pearce, F.R., Couchman,
  H.M.P., ApJ, 552, L27
\bibitem[1997]{ouk97}
  Oukbir, J., \& Blanchard, A. 1997, A\&A, 317, 1
\bibitem[1997]{pea97}
  Peacock, J.A., 1997, MNRAS, 284, 885
\bibitem[1980]{pee80}
  Peebles, P.J.E. 1980, The Large-Scale Structure of the Universe
  (Princeton: Princeton Univ. Press)
\bibitem[2000]{mp00}
  Pierre, M., 2000, Procs. of ``Mining the sky'',
  Joint MPA/ESO/MPE conference, preprint astro-ph/0011166
\bibitem[1974]{ps74}
  Press, W.H., \& Schechter, P., 1974, ApJ, 187, 425
\bibitem[1977]{rs77}
  Raymond, J.C., Smith, B.W., 1977, ApJS, 35, 419
\bibitem[2001]{ric01}
  Ricker, P.M., \& Sarazin, C.L., 2001, ApJ, 561, 621
\bibitem[2000]{rob00}
  Robinson, J. 2000, submitted to MNRAS, astro-ph/0004023
\bibitem[2001]{rom01}
  Romer, A.K., Viana, P.T.P., Liddle, A.R., \& Mann, R.G.,
  ApJ, 547, 594
\bibitem[1998]{sad98}
  Sadat, R., Blanchard, A., \& Oukbir, J., 1998, A\&A,  329, 21
\bibitem[1998]{sp98}
  Starck, J.-L., Pierre, M., 1998, A\&AS 128, 397
\bibitem[2000]{sut00}
  Suto, Y., Yamamoto, K., Kityama, T., \& Jing, Y.P., 2000, ApJ, 534,
  551
\bibitem[2001]{val01}
  Valtchanov, I., Pierre, M. \& Gastaud, R., 2001, A\&A, 370, 689 (VPG)
\bibitem[1996]{via96}
  Viana, P.T.P \& Liddle, A.R., 1996, MNRAS, 281, 323
\bibitem[1999]{via99}
  Viana, P.T.P \& Liddle, A.R., 1999, MNRAS, 303, 535
\bibitem[2001]{wat01}
  Watson, M.G., Augu\`eres, J.-L., Ballet, J., et al. 2001, A\&A, 365,
  L51

\end{thebibliography}
\end{document}